\documentclass[paper]{JHEP3}
\usepackage{cite,axodraw,epsfig}

\newcommand{\be}{\begin{equation}}
\newcommand{\ee}{\end{equation}}
\newcommand{\br}{\begin{eqnarray}}
\newcommand{\er}{\end{eqnarray}}
\newcommand{\ba}{\begin{array}}
\newcommand{\ea}{\end{array}}
\newcommand{\bi}{\begin{itemize}}
\newcommand{\ei}{\end{itemize}}
\newcommand{\bn}{\begin{enumerate}}
\newcommand{\en}{\end{enumerate}}
\newcommand{\bc}{\begin{center}}
\newcommand{\ec}{\end{center}}

\newcommand{\Dir}{\kern -6.4pt\Big{/}}
\newcommand{\Dirin}{\kern -10.4pt\Big{/}\kern 4.4pt}
\newcommand{\DDir}{\kern -8.0pt\Big{/}}
\newcommand{\DGir}{\kern -6.0pt\Big{/}}

\def\frac#1#2{ {{#1} \over {#2} }}

\def\beq{\begin{equation}}
\def\beeq{\begin{eqnarray}}
\def\eeq{\end{equation}}
\def\eeeq{\end{eqnarray}}
\def\a0{\bar\alpha_0}
\def\thrust{\mbox{T}}
\def\Thrust{\mathrm{\tiny T}}

\def\as{\alpha_{\mathrm{S}}}
\def\aem{\alpha_{\mathrm{EM}}}
\def\b0{\beta_0}

\def\ee{e^+e^-}

\def\lms{\Lambda^{(n_{\rm f}=4)}_{\overline{\mathrm{MS}}}}

\def\MSbar{\overline{\mathrm{MS}}}

\def\ycut{y_{\mathrm{cut}}}
\def\slashchar#1{\setbox0=\hbox{$#1$}           
     \dimen0=\wd0                                 
     \setbox1=\hbox{/} \dimen1=\wd1               
     \ifdim\dimen0>\dimen1                        
        \rlap{\hbox to \dimen0{\hfil/\hfil}}      
        #1                                        
     \else                                        
        \rlap{\hbox to \dimen1{\hfil$#1$\hfil}}   
        /                                         
     \fi}                                         %

\def\be{\begin{equation}}
\def\ee{\end{equation}}
\def\bea{\begin{eqnarray}}
\def\eea{\end{eqnarray}}
\def\lsim{\:\raisebox{-0.5ex}{$\stackrel{\textstyle<}{\sim}$}\:}
\def\gsim{\:\raisebox{-0.5ex}{$\stackrel{\textstyle>}{\sim}$}\:}

\def\aem{\alpha_{\rm EM}}

\def\slash{/\kern -5pt}

\def\ims #1 {\ensuremath{M^2_{[#1]}}}

\def\s22w{s_{2W}^2}

\title{Weak Corrections to Three-Jet Production in Electron-Positron 
Annihilations: 1) The Factorisable Contributions\footnote{Work supported in 
part by the U.K.\ Particle Physics and
Astronomy Research Council (PPARC),
by the European Union (EU) under contract HPRN-CT-2000-00149 and by the 
Italian Ministero dell'Istruzione, dell'Universit\`a e della Ricerca
(MIUR) under
contract 2001023713\_006.}}
\author{E. Maina\\
Dipartimento di Fisica Teorica -- Universit\`a di Torino\\
Via Pietro Giuria 1, 10125 Torino, Italy\\
and \\
Istituto Nazionale di Fisica Nucleare -- Sezione di Torino\\
Via Pietro Giuria 1, 10125 Torino, Italy\\
E-mail: {\tt maina@to.infn.it}}
\author{S. Moretti\\
CERN -- Theory Division\\
CH-1211 Geneva 23, Switzerland\\
and\\
Institute for Particle Physics Phenomenology -- University of Durham\\
South Road, Durham DH1 3LE, UK\\
E-mails: {\tt stefano.moretti@cern.ch, stefano.moretti@durham.ac.uk}}
\author{D.A. Ross\\
Department of Physics and Astronomy -- University of Southampton\\
Highfield, Southampton SO17 1BJ, UK\\
Email: {\tt dar@hep.phys.soton.ac.uk}} 
\abstract {We report on the calculation of the
 factorisable one-loop weak-interaction corrections
to  the initial and final states for  
 three-jet observables in electron-positron annihilations.
We show that such corrections are of a few percent at 
$\sqrt s=M_Z$. Hence, while their impact is not dramatic in the context
of LEP1 and SLC, where the total error on the measured value of
$\alpha_{\mathrm{S}}$ is larger, at a future 
Linear Collider, running at the $Z$ mass peak (e.g., GigaZ),
they ought to be taken into account in the experimental fits,
as here the uncertainty on the value of the strong coupling
constant is expected to be at the $0.1\%$ level or even smaller.   
The calculation has been performed using helicity amplitudes so that
it can be applied to the case of polarised beams.}

\preprint{{DFTT 27/02}\\ 
{CERN-TH/2002-191}\\
{IPPP/02/48}\\
{DCPT/02/96}\\
{SHEP-02/24}\\ 
{March 2003}}
\begin{document}
\section{Electroweak corrections at high energies}
\label{Sec:EWCorr}

Strong (QCD) and Electroweak (EW) interactions
are two fundamental forces of 
Nature, the latter in turn unifying
weak and electromagnetic (EM) interactions. 
Together they constitute the Standard Model (SM) of particle physics.
A clear hierarchy exists between the strengths of the two interactions
at the energy scales probed by past and present high energy particle 
accelerators (e.g., LEP, SLC, HERA, RHIC and Tevatron) or, indeed,
at a future generation electron-positron Linear
Collider (LC) \cite{LCs}, running with very high luminosity at
$\sqrt s=M_Z$ (the so-called `GigaZ' stage, where $s$ is the collider CM energy
squared): 
QCD forces are stronger than EW ones. This is quantitatively
 manifest if one recalls that the value of the QCD coupling `constant', 
$\alpha_{\mathrm{S}}$, measured at these machines is much larger 
than the EW one, $\alpha_{\mathrm{EW}}$, typically, by an order of
magnitude.

A peculiar feature distinguishing QCD and EW effects in higher orders
is that the latter are enhanced by double logarithmic factors,
$\log^2(\frac{s}{M^2_{{W}}})$, 
which, unlike in the former, do not cancel for `infrared-safe' 
observables  \cite{Kuroda:1991wn} --\cite{Ciafaloni:1999xg}. 
The origin of these `double logs' is understood.
It is due to a lack of cancellation of infrared (both soft
and collinear) virtual and real emission in
higher order contributions. This is in turn a consequence of the 
violation of the Bloch-Nordsieck theorem in non-Abelian theories
\cite{Ciafaloni:2000df}\footnote{Recently, it has been
found that Bloch-Nordsieck violation can also occur in spontaneously
broken Abelian gauge theories, if the incoming particles are mass
eigenstates that do not coincide with gauge eigenstates
\cite{Ciafaloni:2001vt}. In the SM this is particularly 
relevant for incoming longitudinal gauge bosons or Higgs scalars
\cite{Ciafaloni:2001vu}.}.
The problem is in principle present also in QCD. In practice, however, 
it has no observable consequences, because of the final averaging of the 
colour degrees of freedom of partons, forced by their confinement
into colourless hadrons. This does not occur in the EW case,
where, e.g., the initial state has a non-Abelian charge,
dictated by the given collider beam configuration, such as in $e^+e^-$
collisions. 

These logarithmic corrections are finite (unlike in
QCD), as the masses of the EW gauge bosons provide a physical
cut-off for $W$ and $Z$ emission. Hence, for typical experimental
resolutions, softly and collinearly emitted weak bosons need not be included
in the production cross-section and one can restrict oneself to the 
calculation of weak effects originating from virtual corrections and
affecting a purely hadronic final state. Besides, these contributions can  be
isolated in a gauge-invariant manner from EM effects
\cite{Ciafaloni:1999xg}, \cite{Beccaria:2000fk} --\cite{Beccaria:2001yf},
at least in some simple cases (including the process $e^+e^-\to 
\gamma^*,Z\to$~jets considered here) and therefore may or may not
be included in the calculation, depending on the observable being studied. 
(See Refs.~\cite{Beccaria:2000fk} --\cite{Layssac:2001ur}
for a collection of papers dealing with resummed, one- and two-loop 
EW corrections to various high energy processes.)

\section{One-loop weak effects in three-jets events at leptonic colliders}
\label{Sec:ee}
 
It is the aim of our paper to report on the computation of 
one-loop weak effects entering three-jet production in electron-positron
annihilation at $\sqrt s=M_Z$
via the subprocess $e^+e^-\to\gamma^*,Z\to \bar 
qqg$\footnote{See Ref.~\cite{2jet} for the corresponding weak corrections
to the Born process $e^+e^-\to\bar qq$ and Ref.~\cite{4jet} for the
$\sim n_{\rm f}$ component of those to $e^+e^-\to \bar qqgg$ (where 
$n_{\rm f}$ represents the number of light flavours).
For two-loop results on the former, see \cite{Beenakker:2000kb}.},
when the higher order effects arise only from initial or final state
interactions. 
These represent the so-called `factorisable' corrections, i.e.,
those involving loops 
not connecting the initial leptons to the final quarks,
which are the dominant ones at $\sqrt s=M_Z$ (where the width 
of the $Z$ resonance provides a natural cut-off for off-shellness
effects). The remainder, `non-factorisable' corrections,
while being negligible at $\sqrt s=M_{Z}$, 
are expected to play a quantitatively relevant role as $\sqrt s$ grows
larger.  (The study of the full set
of one-loop weak corrections will be the subject of a future publication.)
As a whole,
one-loop weak effects will become comparable to QCD ones
 at future LCs running at TeV energy scales\footnote{For example, 
at one-loop level,
in the case of the inclusive cross-section of $e^+e^-$
into hadrons, the QCD corrections are of  ${\cal O}
(\frac{\alpha_{\mathrm{S}}}{\pi})$, whereas
the EW ones are of ${\cal O}(\frac{\alpha_{\mathrm{EW}}}{4\pi}\log^2
\frac{s}{M^2_{{W}}})$, where $s$ is the collider CM energy
squared, so that at $\sqrt s=1.5$ TeV the former are identical to the latter,
of order 9\% or so.}. In contrast, 
at the $Z$ mass peak, where no logarithmic
enhancement occurs, one-loop weak effects are expected to appear
at the percent level, hence being of limited relevance at
LEP1 and SLC, where the final error on $\alpha_{\mathrm{S}}$
is of the same order or larger \cite{Dissertori}, but of crucial importance
at a GigaZ stage of a future LC, where the relative accuracy
of $\alpha_{\mathrm{S}}$ measurements is expected to be at the
$0.1\%$ level or smaller \cite{Winter}.
On the subject of higher order QCD
effects, it should be mentioned here that a great deal of effort has    
recently been devoted to evaluate two-loop contributions
to the three-jet process (albeit, only at the amplitude level so far,
as there are no numerical results available yet)
while the one-loop QCD results have been known for some time \cite{ERT}.
Even though a
full ${\cal O}(\alpha_{\mathrm{S}}^3)$ analysis is not yet available, one can
reasonably argue that at $\sqrt s=M_Z$ the two-loop QCD corrections should be 
comparable to the one-loop weak effects computed here. 

In 
the case of $e^+e^-$ annihilations, the most important QCD quantity to be 
extracted from multi-jet events is precisely $\alpha_{\mathrm{S}}$.
The confrontation of the measured value of the strong coupling
constant with that predicted by the theory through the 
renormalisation group evolution is an important test of the
SM or else an indication of new physics, whose typical mass scale is larger
than the collider energy, 
but which can manifest itself through virtual effects. 
Jet-shape observables, which offer a handle on non-perturbative
QCD effects via large power corrections, would be affected as well.

A further aspect that should be recalled is that weak corrections naturally
introduce parity-violating effects in jet observables, detectable through
asymmetries in the cross-section, which are often regarded as an indication
of physics beyond the SM. These effects are further enhanced if polarisation
of the incoming beams is exploited.  
The option of exploiting beam polarisation  is
one of the strengths of the LC projects. 
Comparison of theoretical predictions 
involving parity-violation with future experimental data 
is regarded as another powerful tool for confirming or 
disproving  the existence of some beyond the SM scenarios, such as those 
involving right-handed weak currents and/or new massive gauge bosons.

The plan of the rest of the paper is as follows. In the next Section,
we describe the calculation. Then, in Sect.~\ref{Sec:Results},
we present some numerical results. We conclude in Sect.~\ref{Sec:Conclusions}.

\section{Calculation}
\label{Sec:Calculation}

Since we are considering weak corrections that can be
identified via their induced parity-violating effects and since we wish to
apply our results to the case of polarised electron and/or positron  
beams, it is convenient to work in terms of helicity matrix elements
(MEs). Thus, we define the helicity amplitudes 
${\cal A}^{(G)}_{\lambda_1, \lambda_2, \sigma}$ 
for a gauge boson of type $G$ (hereafter,
a virtual photon $\gamma^*$ or a $Z$-boson) of helicity $\lambda_1$ decaying
into a gluon with helicity $\lambda_2$, a massless quark
with helicity $\sigma$ and a massless
 antiquark with opposite helicity\footnote{Note 
that all interactions considered here preserve the helicity along
the fermion line, including
those in which Goldstone bosons appear inside the loop, since these either
 occur in pairs or involve a mass insertion on the fermion line.}. 
Since the photon is off-shell, it can have a longitudinal polarisation
component, so that the helicity $\lambda_1$ always takes three values,
 $\pm1, \, 0$, for both the $\gamma^*$ and $Z$ 
gauge vectors\footnote{These helicities,
wherein $\pm1(0)$ are(is) transverse(longitudinal), are defined in a 
frame in which the particle
is {\it not} at rest, so that a fourth possible polarisation in the direction
of its four-momentum is irrelevant since its contribution
vanishes by virtue of current conservation.},
whereas $\lambda_2$ and $ \sigma$ can only be equal to $\pm 1$.
 

The general form of these amplitudes may be written as
\be {\cal A}^{(G)}_{\lambda_1, \lambda_2, \sigma}
 \ = \ \bar{u}(p_2) \Gamma \frac{\left(1+\sigma \gamma^5\right)}{2} v(p_1),
\ee
where $p_1$ and $p_2$ are the momenta of the outgoing antiquark and quark 
respectively and $\Gamma$ stands for a sum of strings
 of Dirac $\gamma-$matrices
with coefficients, which, beyond tree level, 
involve integrals over loop momenta.
Since the helicity $\sigma$ of the fermions is conserved the strings 
must contain an odd number of $\gamma-$matrices. Repeated use of the
Chisholm 
identity\footnote{This identity is only valid in four dimensions.
In our case, where we do not have infrared (i.e., 
soft and collinear) divergences, 
it is a simple matter to isolate the ultraviolet divergent contributions,
which are proportional to the tree-level MEs,
 and handle them separately. However, in $d$ dimensions
one needs to account for the fact that there are $2^{d/4}$ 
helicity states for the fermions and $(d-2)$ for the gauge bosons.
The method described here will {not} correctly trap terms
proportional to $(d-4)$ in coefficients of divergent integrals. 
It is probably for this reason that the formalism of Ref.~\cite{twol3p}
is considerably more cumbersome than that presented here.}
 means that $\Gamma$ can always be expressed in the
form
\be \Gamma \ = \ C_1 \, \gamma \cdot p_1 \ + \ 
C_2 \, \gamma \cdot p_2 \ + \ C_3 \, \gamma \cdot p_3 \ + \ 
C_4 \, \sqrt{Q^2} \, \gamma \cdot n , \label{helicityme1} \ee
where $p_3$ is the momentum of the outgoing gluon, $Q^2=(p_1+p_2+p_3)^2$ is
 the square momentum of the gauge boson,
and  $n$ is a unit vector normal to the momenta of the jets, more
precisely:
\begin{equation}
n_\mu =\frac{1}{\sqrt{2~p_1\cdot p_2~p_1\cdot p_3~p_2\cdot p_3}}
\varepsilon_{\mu\nu\rho\sigma}p_1^\nu p_2^\rho p_3^\sigma.
\end{equation}
The coefficient functions $C_i$ depend on the helicities  
$\lambda_1, \ \lambda_2, \ \sigma$ as well as the 
energy fractions $x_1$ and $x_2$ of the antiquark and quark in the final 
state, i.e.,
\begin{equation}
x_i=\frac{2 E_i}{\sqrt s} \qquad (i=1,2),
\end{equation}
 and on all the couplings and masses of particles that enter into
the relevant perturbative contribution to the amplitude. 

For massless fermions the MEs of the first two terms
of eq.~(\ref{helicityme1})
vanish, and we are left with
\begin{eqnarray} \ {\cal A}^{(G)}_{\lambda_1, \lambda_2, \sigma} &=& 
C_3 \, \bar{u}(p_2) \gamma \cdot p_3 \frac{\left(1+\sigma \gamma^5\right)}{2}
 v(p_1) \, + \, C_4 \sqrt{Q^2}
\, \bar{u}(p_2) \gamma \cdot n \frac{\left(1+\sigma \gamma^5\right)}{2}
 v(p_1), \nonumber \\ &=& 
C_3 \, Q^2 \sqrt{(1-x_1)(1-x_2)} \ - \ i \, \sigma\, C_4 \, Q^2 
  \sqrt{x_1+x_2-1}. \end{eqnarray}

The relevant coefficient functions $C_3$ and $C_4$ 
are scalar quantities and can be projected on a graph-by-graph basis
using the projections
\be C_3 \ = \ {\rm Tr} \left( \Gamma \gamma \cdot v
\frac{\left(1+\sigma \gamma^5\right)}{2}
 \right),\ee  
where $v$ is the vector
$$ v \ = \ \frac{(1-x_2) p_1 + (1-x_1) p_2 - (x_1+x_2-1) p_3}
{2 Q^2 (1-x_1)(1-x_2)},
$$
and
\be C_4 \ = \ -\frac{1}{2\sqrt{Q^2}}{\rm Tr} \left( \Gamma \gamma \cdot n
\frac{\left(1+\sigma \gamma^5\right)}{2}
 \right)  .\ee

At tree level the helicity amplitudes are only functions of $x_1, \ x_2$,
 the EW couplings $g_j^{(G)}$ of the (anti)quark of type $j$
(proportional to $g_{\rm{W}}\equiv\sqrt {4\pi \alpha_{\rm{EW}}}$,
with $\alpha_{\rm{EW}}=\alpha_{\rm{EM}}/\sin^2\theta_W$, and
carrying information on both helicity {and} flavour
of the latter) to the relevant
gauge boson and the QCD coupling $g_{\rm S}\equiv
\sqrt{4\pi\alpha_{\mathrm S}}$. Specifically, in case of massless
(anti)quarks (i.e., $m_q=0$), we have (here, $\tau^a$ represents a colour
matrix):
\br\label{LO}
{\cal A}^{(G)}_{1,1,1} \ = \ {\cal A}^{(G)}_{-1,-1,-1}
 & = & -2 i g_j^{(G)} g_{\mathrm S} \tau^a \frac{x_1}{\sqrt{(1-x_1)(1-x_2)}},
\nonumber \\
{\cal A}^{(G)}_{1,1,-1} \ = \ {\cal A}^{(G)}_{-1,-1,1}
 &=& -2 i g_j^{(G)} g_{\mathrm S} \tau^a \frac{\sqrt{(1-x_1)(1-x_2)}}{x_1},
\nonumber \\
{\cal A}^{(G)}_{0,-1,1} \ = \ {\cal A}^{(G)}_{0,1,-1}
 &=& - 2 \sqrt{2} i g_j^{(G)} g_{\mathrm S} \tau^a {\sqrt\frac{(1-x_1-x_2)}{x_1}},
\er
with all others being zero. These zero values do {not}, in general,
remain zero in the presence of weak corrections 
and this can lead to a relative enhancement of the latter, in comparison
to QCD effects at the same order.

\begin{figure}
\begin{center}
\begin{picture}(365,225)
\Photon(0,175)(50,175){4}{4}
\ArrowLine(150,125)(125,137) \Line(125,137)(100,150)
\ArrowLine(100,150)(75,162) \ArrowLine(75,162)(50,175)
\ArrowLine(50,175)(150,225) \Gluon(115,142)(150,150){4}{3}
\PhotonArc(87,155)(12,337,150){3}{5} \put(87,175){$W,Z$}
\put(10,185){$Z,\gamma$}

\Photon(200,175)(250,175){4}{4}
\ArrowLine(350,125)(325,137) \ArrowLine(325,137)(300,150)
\ArrowLine(300,150)(275,162) \ArrowLine(275,162)(250,175)
\ArrowLine(250,175)(350,225) \Gluon(275,162)(350,175){4}{8}
\PhotonArc(312,143)(12,334,150){3}{5} \put(324,152){$W,Z$}
\put(210,185){$Z,\gamma$}

\Photon(0,50)(50,50){4}{4}
\ArrowLine(150,0)(100,25) \ArrowLine(100,25)(50,50)
\ArrowLine(50,50)(90,72) \ArrowLine(90,72)(110,82)
\ArrowLine(110,82)(150,100) \Gluon(100,25)(150,35){4}{5}
\PhotonArc(100,76)(12,32,206){3}{5} \put(87,95){$W,Z$}
\put(10,60){$Z,\gamma$}

\Photon(200,50)(240,50){4}{4} \Photon(265,50)(300,50){4}{4}
\ArrowLine(350,0)(325,25) \ArrowLine(325,25)(300,50)
\ArrowLine(300,50)(350,100) \Gluon(321,29)(350,45){4}{3}
\GCirc(252,50){12}{.5}
\put(210,60){$Z,\gamma$}

\end{picture}
\vspace{1.0truecm}
\caption{Self-energy insertion graphs. The shaded blob on the incoming
wavy line represents all the contributions to the gauge boson 
self-energy and is dependent on the Higgs mass (hereafter, we will use
$M_H=115$ GeV for the latter). In this and all subsequent
figures the graphs in which the exchanged gauge boson is a $W$-boson
is accompanied by corresponding graphs in which the  $W$-boson is replaced
by its corresponding Goldstone boson. Since the Yukawa couplings are
proportional to the fermion masses, such graphs are only significant
 in the case of $b$-quark jets. There is a similar set of diagrams
in which the direction of the fermion line is reversed.}
\label{se_graphs}
\end{center}
\end{figure}
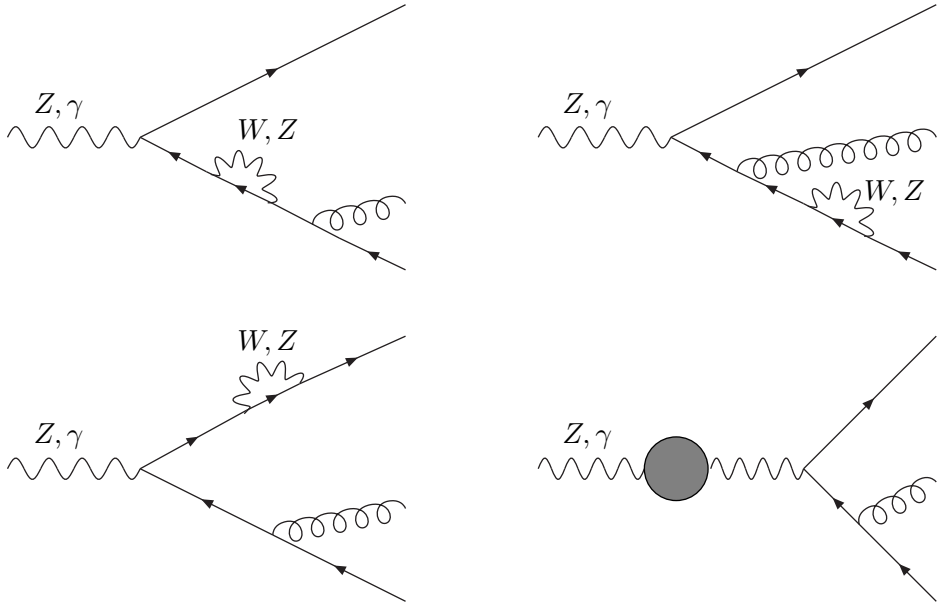

\begin{figure}
\begin{center}
\begin{picture}(365,225)

\Photon(100,175)(150,175){4}{4}
\ArrowLine(250,125)(225,137) \ArrowLine(225,137)(200,150)
\ArrowLine(200,150)(175,162) \ArrowLine(175,162)(150,175)
\ArrowLine(150,175)(250,225) \Gluon(200,150)(250,170){4}{5}
\PhotonArc(200,150)(24,154,330){3}{8} \put(187,115){$W,Z$}
\put(110,185){$Z,\gamma$}

\Photon(0,50)(50,50){4}{4}
\ArrowLine(150,0)(125,12) \ArrowLine(125,12)(87,30)
 \ArrowLine(87,30)(50,50)
\ArrowLine(50,50)(87,70) \ArrowLine(87,70)(150,100)
 \Gluon(115,17)(150,25){4}{3}
\Photon(97,25)(97,75){3}{5} \put(102,50){$W,Z$}
\put(10,60){$Z,\gamma$}

\Photon(200,50)(250,50){4}{4}
\ArrowLine(350,0)(325,12) \ArrowLine(325,12)(297,25)
 \Photon(297,25)(250,50){3}{5}
\Photon(250,50)(297,75){-3}{5} \ArrowLine(297,75)(350,100)
 \Gluon(315,17)(350,25){4}{3}
\ArrowLine(297,25)(297,75) \put(275,73){$W$} \put(275,19){$W$}
\put(210,60){$Z,\gamma$}

\end{picture}
\vspace{1.0truecm}
\caption{Vertex correction  graphs.  Again, same considerations 
as in the previous figure apply for the case of Goldstone bosons and
there is a similar set of graphs
in which the direction of the fermion line is reversed} \label{vertex_graphs}
\end{center}
\end{figure}
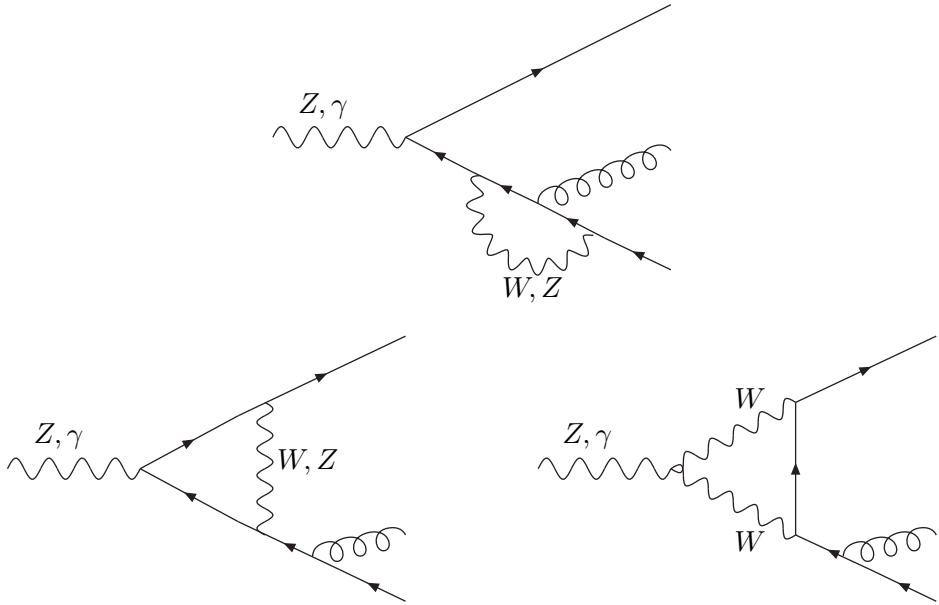

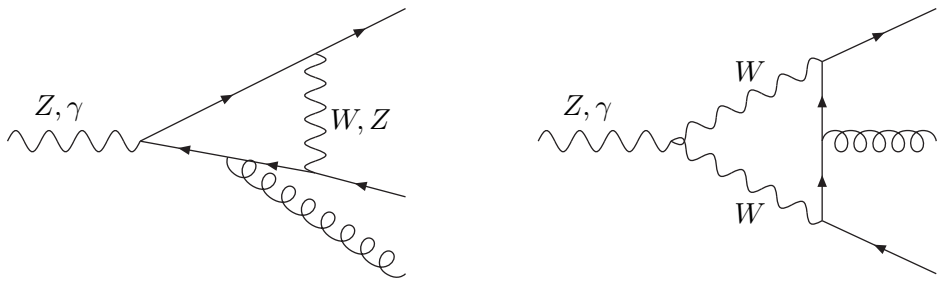
\begin{figure}
\begin{center}
\begin{picture}(365,105)

\Photon(0,50)(50,50){4}{4}
\put(10,60){$Z,\gamma$}
\ArrowLine(150,29)(116,38) \ArrowLine(116,38)(83,44) \ArrowLine(83,44)(50,50)
\ArrowLine(50,50)(116,83) \ArrowLine(116,83)(150,100)
\Gluon(83,44)(150,0){-4}{8} \Photon(116,38)(116,83){4}{5}
\put(122,55){$W,Z$}

\Photon(200,50)(250,50){4}{4}
\ArrowLine(350,0)(307,20) 
 \Photon(307,20)(250,50){3}{5}
\Photon(250,50)(307,80){-3}{5} \ArrowLine(307,80)(350,100)
 \Gluon(307,50)(350,50){4}{5}
\ArrowLine(307,20)(307,50) \ArrowLine(307,50)(307,80)
 \put(275,73){$W$} \put(275,19){$W$}
\put(210,60){$Z,\gamma$}

\end{picture}
\vspace{1.0truecm}
\caption{Box  graphs. Again, same considerations as in the previous 
two figures apply for the case of Goldstone bosons. Here, the first 
graph is accompanied by a similar graph with the direction of the 
fermion line reversed whereas for the second
graph this reversal does {not} lead to a distinct
Feynman diagram.} \label{box_graphs}
\end{center}
\end{figure}

At one-loop level such helicity amplitudes   
acquire higher order corrections from the self-energy insertions on the
fermions and gauge bosons shown in Fig.~\ref{se_graphs},
from the vertex corrections shown in Fig.~\ref{vertex_graphs}\footnote{Note
that we also include self-energy and vertex corrections to the incoming
$e^+e^-\to \gamma^*,Z$ current, though we do not show the corresponding
graphs.}
and from the box diagrams shown in Fig.~\ref{box_graphs}. As we have neglected 
here the masses of the final-state quarks, such higher order corrections
depend on the ratio $Q^2/M_{{W}}^2$, where $Q^2$ is the square
momentum of the gauge boson, as well as the EM coupling
constant $\alpha_{\rm EM}$ and the weak mixing angle $s_W\equiv\sin\theta_W$
(with $\alpha_{\rm EW}=\alpha_{\rm EM}/s_W^2$).
Furthermore, in the case where the final state fermions are $b$-quarks, 
the loops involving the exchange of a $W$-boson lead to effects of
virtual $t$-quarks, so that the corrections also depend on 
the ratio $m_t^2/M_{{W}}^2$. (It is only in this case that the graphs 
involving
the exchange of the Goldstone bosons associated with the $W$-boson graphs
are relevant.)

The self-energy and vertex correction graphs contain ultraviolet divergences.
These have been subtracted using the `modified' Minimal Subtraction
($\MSbar$) scheme at
the scale $\mu=M_Z$. Thus the couplings are taken to be
those relevant for such a subtraction: e.g., the EM coupling,
$\alpha_{\mathrm{EM}}$, has been taken to be $1/128$ at the above subtraction
point. 
Two exceptions to this
renormalisation scheme have been the following:
\begin{enumerate}
\item the self-energy insertions
on external fermion lines, which have been subtracted on mass-shell,
so that the external fermion fields create or destroy particle states
with the correct normalisation;
\item the mass renormalization of the $Z$-boson propagator, which has also been 
carried out on mass-shell, so that the $Z$ mass does indeed refer to the
physical pole-mass.
\end{enumerate}

All these graphs are infrared and collinear convergent so that they
 may be expressed in terms of Passarino-Veltman \cite{VP} functions
which are then evaluated numerically. The expressions for
 each of these diagrams 
have been calculated using FORM \cite{FORM} and checked by an
independent program based on FeynCalc \cite{FeynCalc}. For the numerical
evaluation of the scalar integrals we have relied on FF \cite{FF1.9}. 
A further check on our results has been carried out
by setting the polarisation vector of the photon proportional to its momentum
and verifying that in that case the sum of all one-loop diagrams
vanishes, as required by gauge invariance.
The full expressions for the contributions from these graphs are too
lengthy to be reproduced here.

In terms of the helicity MEs we define the following 
``spin-matrix'' tensors, only depending on the 
polarisation state of the off-shell gauge boson,
\be {\cal T}^{(GG^\prime)}_{\lambda\lambda^\prime}  \ = \ 
 \sum_{\lambda_2,\sigma} 
{\cal A}^{(G)}_{\lambda, \lambda_2, \sigma}
\left({\cal A}^{(G^\prime)}_{\lambda^\prime, \lambda_2, \sigma}\right)^\dagger,
\ee
where the (anti)quark and gluon helicities have been summed over.
These tensor elements are real at tree level, but in general acquire an
imaginary part at one loop arising from the cuts of the
loop integrations above the threshold for the production
of the internal particles. 

Finally, we define the customary nine form-factors, $F_1, ... F_9$,
describing the differential structure of a three-jet final state
in terms of the above spin-matrix tensors, as follows:
\br\label{FFs}
 F_i \ & = & \ \frac{\alpha_{\mathrm{EM}}}{512\pi^3} \left[
\frac{\left(\eta_A^{L(R)}\right)^2}{Q^2} f_i^{AA} 
\right. \nonumber \\ & & \left. +
 \frac{(1-\lambda_e-4 s_W^2)}{4s_W \, c_W}
 \eta_A^{L(R)} \eta_Z^{L(R)} 
  \left(f_i^{AZ}+f_i^{ZA}\right)
   \Re e \left\{ \frac{1}{\left( Q^2-M_Z^2+i\Gamma_Z M_Z \right)} \right\}
\right. \nonumber \\ & & \left.  \hspace*{-1.25cm} +
 \left(\frac{1-\lambda_e-4s_W^2)}{4s_W \, c_W}\right)^2
Q^2 (\eta_Z^{L(R)})^2  f_i^{ZZ}
\Re e \left\{ \frac{1}{ \left(Q^2-M_Z^2+i\Gamma_Z M_Z \right)^2} \right\} 
 \right]
\ \  (i=1, ... 9), \label{eq33}
\er
where
\br f_1^{GG^\prime} &=& \left( {\cal T}^{GG^\prime}_{1,1}
 +{\cal T}^{GG^\prime}_{0,0} +{\cal T}^{GG^\prime}_{-1,-1} \right), \nonumber \\
 f_2^{GG^\prime} &=& {\cal T}^{GG^\prime}_{0,0}, \nonumber \\
 f_3^{GG^\prime} &=& -2 \, \Re e  \left( {\cal T}^{GG^\prime}_{1,1}
 -{\cal T}^{GG^\prime}_{-1,-1} \right), \nonumber \\
 f_4^{GG^\prime} &=& -\sqrt{2} \, \Re e \left( {\cal T}^{GG^\prime}_{1,0}
 +{\cal T}^{GG^\prime}_{-1,0} \right), \nonumber \\
 f_5^{GG^\prime} &=& -2 \, \Re e  {\cal T}^{GG^\prime}_{1,-1}, \nonumber \\
 f_6^{GG^\prime} &=& 2 \sqrt{2}\, \Re e \left( {\cal T}^{GG^\prime}_{1,0}
 - {\cal T}^{GG^\prime}_{-1,0} \right), \nonumber \\
 f_7^{GG^\prime} &=& \sqrt{2} \, \Im m\left( {\cal T}^{GG^\prime}_{0,1}
  -{\cal T}^{GG^\prime}_{0,-1} \right), \nonumber \\
 f_8^{GG^\prime} &=& 2 \, \Im m {\cal T}^{GG^\prime}_{1,-1}, \nonumber \\
 f_9^{GG^\prime} &=& -2 \sqrt{2} \, \Im m \left( {\cal T}^{GG^\prime}_{0,1}
 +{\cal T}^{GG^\prime}_{0,-1} \right), \label{eq34}
\er
with $\eta_G^{L(R)}$ the weak correction factor to the coupling of the
left(right)-handed electron to the gauge boson $G$ and $\lambda_e$
the helicity of the incoming electron beam (assumed always to be
of opposite helicity to the incoming positron beam).

Up to an overall constant, these form-factors are the same as those 
introduced, e.g., in Refs. \cite{BDS,KS}.
The last three ($F_7, ... F_9$) can arise for the first time at the one-loop 
level, since they are proportional to the imaginary parts of the spin-matrix.
Besides, $F_3, \ F_6$ and $F_7$ vanish in the parity-conserving limit and 
can therefore be used
as probes of weak interaction contributions to three-jet production.
(Moreover, $F_3$ and $F_6$ would be exactly 
 zero at tree level if the leading order 
process were only mediated by virtual photons.)

These form-factors further generate the double differential cross-section for 
three-jet production in terms of some event shape variable, $S$, 
which is in turn related to  $x_1, \, x_2$ by some function, $s$, i.e.,
$S = s(x_1,x_2) $,
and of the polar and azimuthal angles, $\alpha, \, \beta$, 
between, e.g., the incoming electron beam and the antiquark jet, 
by\footnote{A qualitative difference between the expressions
of the form-factors, $F_i$ ($i=1,...9$),
 used here and those of Refs. \cite{BDS,KS} is that we do {not}
include the sign of the axial vector coupling of the electron
to the exchanged gauge boson in our definitions.
 In this way the difference between the differential cross-sections
for left- and right-handed polarised electron beams is manifest
 in eq.~(\ref{eq35}).}
\br\label{angles}
 \frac{d^3\sigma}{dS \, d\cos\alpha \, d\beta} & =  &\int dx_1 dx_2
 ~\delta \left(S-s(x_1,x_2) \right)    
\left[ (2-\sin^2\alpha) \, F_1  +(1-3\cos^2\alpha) 
\, F_2
\right. \nonumber \\ 
 & &  \hspace*{1cm}  + \, \lambda_e \,
 \cos\alpha \, F_3  + \sin 2\alpha \cos\beta \, F_4 
+ \sin^2\alpha \cos 2\beta \, F_5 + \lambda_e \,
 \sin\alpha \cos\beta \, F_6 \nonumber \\
& & \hspace*{1 cm} + \, \sin 2 \alpha \sin \beta \, F_7 + \sin^2 \alpha \sin 2 \beta \,
 F_8 
\left. + \, \lambda_e \, \sin\alpha \sin\beta \, F_9 \right] . \label{eq35}
 \er  
Note that upon integrating over the antiquark jet angle 
relative to the electron
beam, only the form-factor $F_1$ survives.

In general, it is not possible to distinguish between quark, antiquark
and gluon jets, although the above expression can easily be adapted such that 
the angles $\alpha, \beta$ refer to the leading jet. However, (anti)quark
jets {can} be recognised when they originate from primary 
$b$-(anti)quarks, thanks to rather efficient flavour tagging techniques
(such as  $\mu$-vertex devices). We will therefore consider 
the numerical results for such a case separately.

\section{Numerical results}
\label{Sec:Results}

The processes considered here are the following: 
\begin{equation}\label{procj}
e^+e^-\to \gamma^*,Z\to \bar qqg\quad{\mathrm{(all~flavours)}},
\end{equation}
when no assumption is made on the flavour content of the final state,
so that a summation will be performed over $q=d,u,c,s,b$-quarks, and also
\begin{equation}\label{procb}
e^+e^-\to \gamma^*,Z\to \bar bbg,
\end{equation}
limited to the case of bottom quarks only in the final state. 
As already intimated, all quarks
in the final state of (\ref{procj})--(\ref{procb}) are taken as 
massless\footnote{Mass effects in $e^+e^-\to \gamma^*,Z\to \bar bbg$
have been studied in \cite{BMM} and \cite{bbgNLO}.}.
In contrast, the top quark entering the loops in both reactions has
been assumed to have the mass $m_t=175$ GeV. The $Z$ mass used was
$M_Z=91.19$ GeV and was related to the $W$-mass, $M_W$, via the
SM formula $M_W=M_Z\cos\theta_W$, where $\sin^2\theta_W=0.232$.
(Corresponding widths were $\Gamma_Z=2.5$ GeV and $\Gamma_W=2.08$ GeV.)
For $\alpha_{\rm S}$ we have used the two-loop expression for
$\lms=200$ MeV throughout
(yielding, $\alpha_{\rm S}(M_Z)=0.11$).  

We systematically neglect higher
order effects from EM radiation, including those due to Initial State 
Radiation (ISR) or beamstrahlung. In fact, although these are known 
to be non-negligible
(especially at LC energies), we expect them to have a similar effect on 
both the tree-level and one-loop descriptions, 
hence being irrelevant for our purpose. In this context, we should like
to elaborate further on the purely EM corrections to the final state of
processes (\ref{procj})--(\ref{procb}). 
Those to the form-factor  $F_1$ have already been calculated,
since they can be extracted from the Abelian part of the NLO-QCD corrections
(see \cite{ERT}) by replacing $C_F$ by unity and $\alpha_{\rm S}$ by 
$\alpha_{\rm EM}$.
As was pointed out in Ref.~\cite{DAR}, these corrections are dominated
by a term $\sim\alpha_{\mathrm{EM}} \pi/2$ 
multiplying the tree-level cross-section. This contribution is $\sim 1 \%$
and is independent of the jet event shape. A further correction,
associated with the Sudakov form-factor, acts in the negative direction
and is subdominant  away from the two-jet region (i.e., up to 
values of $\sim0.95$ for the Thrust, see below for its definition).
There is no reason to believe that 
these EM corrections would be enhanced for other form-factors.

It is common in the specialised literature to define the $n$-jet fraction
$R_n(y)$ as
\beq
\label{fn}
R_n(y)=\frac{\sigma_n(y)}{\sigma_0},
\eeq
where $y$  is a suitable
variable quantifying the space-time separation among hadronic objects
and with $\sigma_{0}$ identifying the (energy-dependent) 
Born cross-section for $e^+e^-\to \bar qq$.

For the choice $\mu=\sqrt s$ of the renormalisation scale, 
one can conveniently write the three-jet fraction in the following form:
\beq
\label{f3}
R_3(y) =     \left( \frac{\as}{2\pi} \right)    A(y)
           + \left( \frac{\as}{2\pi} \right)^2  B(y) + ... ,
\eeq
where the coupling constant $\as$ and the functions $A(y)$ and $B(y)$ 
are defined in  the $\overline{\mbox{MS}}$ scheme. An experimental fit
of the $R_n(y)$ jet fractions to the corresponding 
theoretical prediction is a powerful
way of determining $\as$ from multi-jet rates.

Through order ${\cal O}(\alpha_{\rm{S}})$
processes (\ref{procj})--(\ref{procb}) are the leading order (LO) perturbative 
contributions to the corresponding three-jet cross-section\footnote{Hereafter,
perturbative contributions are refereed to relatively to the 
${\cal O}(\alpha_{\rm{EM}}^2)$ two-jet rate.}, as defined via
eqs.~(\ref{fn})--(\ref{f3}). 
The LO terms, however, receive  higher order corrections from both QCD
and EW interactions and we are concerned here with the
next-to-leading order (NLO) ones only. 
Whereas at LO all the contributions to the three-jet cross-section
come from the tree-level parton process $e^+e^-\rightarrow \bar qq g$
(which contributes to the $A(y)$ function above),
at NLO the QCD contributions to the three-jet rate (hereafter, denoted
by NLO-QCD) are due to two sources. First, the real emission diagrams for the 
processes $e^+e^-\rightarrow q\bar q gg$ and 
$e^+e^-\rightarrow q\bar q Q\bar Q$, in which
one of the partons is `unresolved'. This can happen when one has either
two collinear partons within one jet or one soft parton outside the
jet. Both these contributions are (in general, positively) divergent. 
Thanks to the Bloch-Nordsieck \cite{BN} and Kinoshita-Lee-Nauenberg
\cite{KLN} theorems, these collinear
and soft singularities are cancelled at the same order in $\as$  by the 
divergent contributions (generally negative) provided by the second source,
namely, the virtual loop graphs. Therefore, after renormalising
the coupling constant $\as$, a finite three-jet cross-section is obtained and
the function $B(y)$ accounts for the above-mentioned three- and four-parton 
QCD contributions\footnote{In order to calculate these,
we make use here of a program based on Ref.~\cite{EERAD}.}. 
While the EM component of the EW corrections may be treated
on the same footing as the QCD one (with the additional photon playing
the role of a second gluon), the weak corrections of interest (hereafter,
labelled as NLO-W) only contribute to three-parton final states. Hence,
in order to account for the latter, it will suffice to make the replacement
\beq
\label{f3EW}
A(y)\to A(y)+A_{\mathrm{W}}(y)
\eeq
in eq.~(\ref{f3}).

The decision as to whether two hadronic 
objects are unresolved or otherwise is usually taken
through the application onto the hadronic final state
of a so-called `jet clustering algorithm',
wherein the number of {\sl clusters}\footnote{Here
  and in the following, the word `cluster' refers to hadrons or
  calorimeter cells in the real experimental case, to partons in the
  theoretical perturbative calculations and also to intermediate jets
  during the clustering procedure.}  
is reduced one at a time by combining the two most (in some sense)
nearby ones. The joining procedure is stopped by testing against some
criterion and the final clusters are called jets. 

As jet clustering schemes\footnote{We 
acknowledge here the well admitted abuse of referring to the various
jet `finders'
both as algorithms and as schemes, since the last term was originally
intended to identify the composition law of four-momenta when pairing two
clusters: in our case, $p^\mu_{ij}=p^\mu_{i}+p^\mu_{j}$.}, 
we have used a selection of
the `binary' ones, in which only two objects are clustered together at any 
step. Given two clusters
labelled as $i$ and $j$, the measure of their `distance' is normally
denoted by $y_{ij}$ and the minimal separation allowed by $y_{\rm{cut}}$.
The algorithms are the following:
the JADE (J) one \cite{jade}, which uses as a measure of separation 
the quantity
\begin{equation}\label{J}
y_{ij}^{\rm J} = {{2 E_i E_j (1-\cos\theta_{ij})}\over{s}};
\end{equation}
the Durham (D) \cite{durham} and the Cambridge (C)
\cite{cambridge}  ones, both using\footnote{The Cambridge algorithm in 
fact only modifies the clustering procedure of the Durham jet finder and the
two implementations coincide for $n\le 3$ parton final states.}
\begin{equation}\label{D_C}
y_{ij}^{\rm D} \equiv y_{ij}^{\rm C} 
= {{2\min (E^2_i, E^2_j)(1-\cos\theta_{ij})}\over{s}};
\end{equation}
the Geneva (G) one \cite{BKSS},  for which one has
\begin{equation}\label{G}
y_{ij}^{\rm G} = \frac{8}{9} \frac{E_i E_j (1 - \cos\theta_{ij})}%
{(E_i + E_j)^2}.
\end{equation}

In eqs.~(\ref{J})--(\ref{G}), $E_i$ and $E_j$ are the energies
 and $\theta_{ij}$ the angular separation
of any pair $ij$ of clusters in the final state. 
The choice of these particular schemes has a simple motivation.
The D and C ones are different versions of `transverse-momentum'
based algorithms, whereas the J and G ones use an `invariant-mass' measure
(see \cite{schemes} for a review).
In fact, these two categories are those that have so far been employed 
most in phenomenological studies of jet physics
in electron-positron collisions, with the former
gradually overshadowing the latter, thanks to their reduced
scale dependence in higher order QCD (e.g., in the case of the
${\cal{O}}(\as^2)$ three- \cite{schemes} --\cite{mb}
and ${\cal{O}}(\as^3)$ four-jet rates 
\cite{as4}) and to smaller hadronisation effects in the same contexts
(see Refs.~\cite{schemes,BKSS}).

\begin{figure}
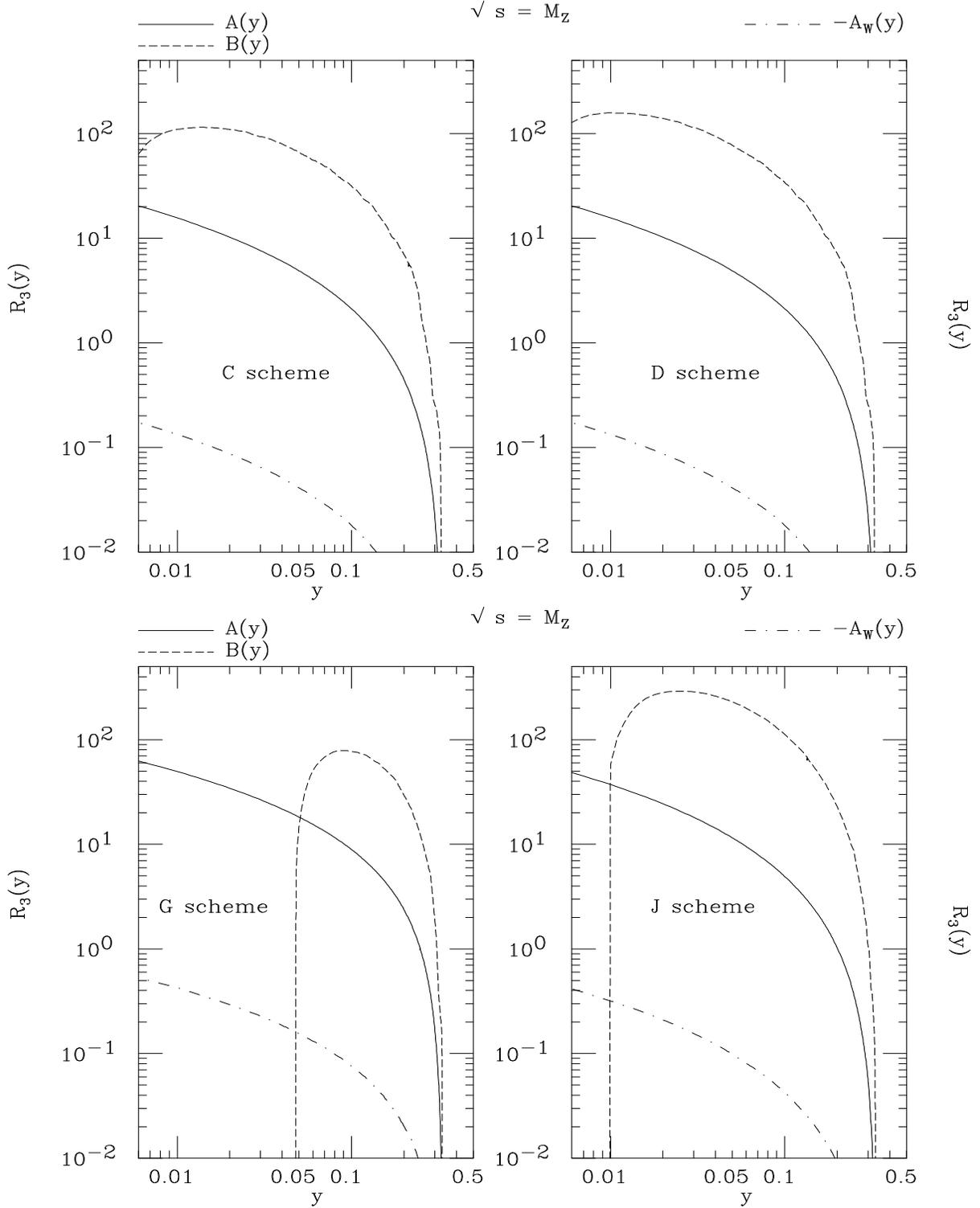

\begin{center}
\epsfig{file=yCD_LEP1_new.ps,height=160mm, width=100mm, angle=90}
\epsfig{file=yGJ_LEP1_new.ps,height=160mm, width=100mm, angle=90}
\end{center}
\vskip -0.5cm
\caption{The $A(y)$, $-A_{\mathrm{W}}$ and $B(y)$ coefficient functions
of eqs.~(\ref{f3})--(\ref{f3EW}) for the Cambridge, Durham, Geneva and
Jade jet clustering algorithms, at $\sqrt s=M_Z$. (Notice that
the $\sim A_{\mathrm{W}}$ term
has been plotted with opposite sign
for better presentation.)}
\label{fig:y_LEP1}
\end{figure}

Fig.~\ref{fig:y_LEP1} displays the $A(y)$, $-A_{\rm{W}}(y)$ 
and $B(y)$ coefficients 
entering eqs.~(\ref{f3})--(\ref{f3EW}), as a function of 
$y(\equiv y_{\rm{cut}})$ for the four above
jet algorithms at $\sqrt s=M_Z$\footnote{Notice 
that $A(y)$ and $A_{\rm{W}}(y)$ for the
C scheme are identical to those for the D one (recall the previous footnote).}.
 A comparison
between $A(y)$ and $A_{\rm{W}}(y)$ reveals that the NLO-W corrections are
negative and remain
indeed at the percent level, i.e., of order $\frac{\aem}{2\pi s_W^2}$
without any logarithmic enhancement (since $\sqrt s\approx M_{{W}}, M_Z$).
They give rise to corrections to $\sigma_3(y)$ of --1\%, and
thus are generally much smaller than the NLO-QCD ones. In this context, no
systematic difference is seen with respect to the choice of jet clustering
algorithm, over the typical range of application of the latter at $\sqrt
s=M_Z$ (say, 
$\ycut\gsim0.005$ for D, C and $\ycut\gsim0.01$ for G, J).

\begin{figure}
\begin{center}
\hskip -0.75cm\epsfig{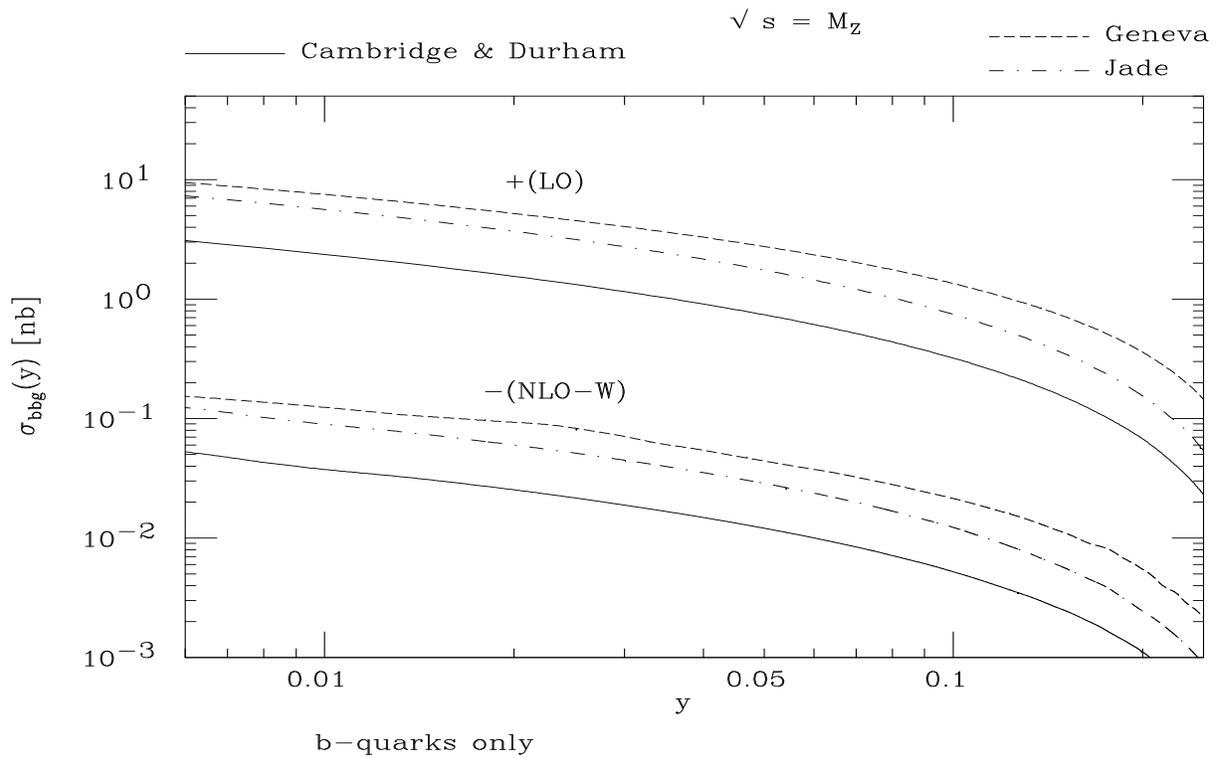}
\end{center}
\vskip -0.5cm
\caption{The total cross section for process (\ref{procb}) at LO
and NLO-W for the Cambridge, Durham, Geneva and
Jade jet clustering algorithms, at $\sqrt s=M_Z$. (Notice that the
NLO-W results have been plotted with opposite sign
for better presentation.)}
\label{fig:y_LEP1_b}
\end{figure}

As already mentioned, it should
now be recalled that jets originating from $b$-quarks can efficiently be
distinguished from light-quark jets. 
Besides, the $b$-quark component of the full three-jet sample is the
only one sensitive to $t$-quark loops in all diagrams of 
Figs.~\ref{se_graphs}--\ref{box_graphs}, hence one may expect somewhat
different effects from weak corrections to process (\ref{procb})
than to (\ref{procj}) (the residual dependence on the $Z \bar q q$
couplings is also different). This is confirmed by 
Fig.~\ref{fig:y_LEP1_b}, where we present the total cross section at $\sqrt
s=M_Z$ for $e^+e^-\to\gamma^*,Z\to\bar bbg$ as obtained at LO and NLO-W, for
our usual choice of jet clustering algorithms and separations. A close
inspection of the plots reveals that NLO-W effects can reach the 
$\sim -2.0\%$ level or so.

In view of these percent effects being well above the error estimate
expected at a future high-luminosity LC running at the $Z$ pole,
it is then worthwhile to further consider the effects
of NLO-W corrections to some other `infrared-safe' jet observables typically
used in the determination
of $\as$, the so-called `shape variables' \cite{KunsztNason}. 
A representative quantity in this respect is the Thrust (T)
distribution \cite{thrust}. This is defined as the sum of 
the longitudinal momenta relative to the (Thrust) axis $n_{\rm T}$ chosen
to maximise this sum, i.e.:
\begin{equation}\label{thrust}
\thrust = {\rm max} \frac{\sum_i |\vec{p_i}\cdot\vec{n_{\mathrm{T}}}|}
                         {\sum_i |\vec{p_i}|} ,
\end{equation} 
where $i$ runs over all final state clusters.
This quantity is identically one at Born level, getting
the first non-trivial
contribution through ${\cal O}(\as)$ from events of the
type (\ref{procj})--(\ref{procb}). Also notice that any other higher 
order contribution will affect this observable. Through ${\cal O}(\as^2)$,
for the choice $\mu =\sqrt s$ of the renormalisation scale, 
the T distribution can be parametrised in the following form:
\begin{equation}\label{T}
(1-{\rm{T}})\frac{d\sigma}{d\thrust}\frac{1}{\sigma_0} = 
\left(\frac{\as}{2\pi}\right)   A^{\Thrust}(\thrust)+
\left(\frac{\as}{2\pi}\right)^2 B^{\Thrust}(\thrust).
\end{equation} 
Again, the replacement 
\begin{equation}\label{TW}
A^{\Thrust}(\thrust)\to A^{\Thrust}(\thrust)+A^{\Thrust}_{\rm{W}}(\thrust) 
\end{equation} 
accounts for the inclusion of the NLO-W contributions.

\begin{figure}
\begin{center}
\epsfig{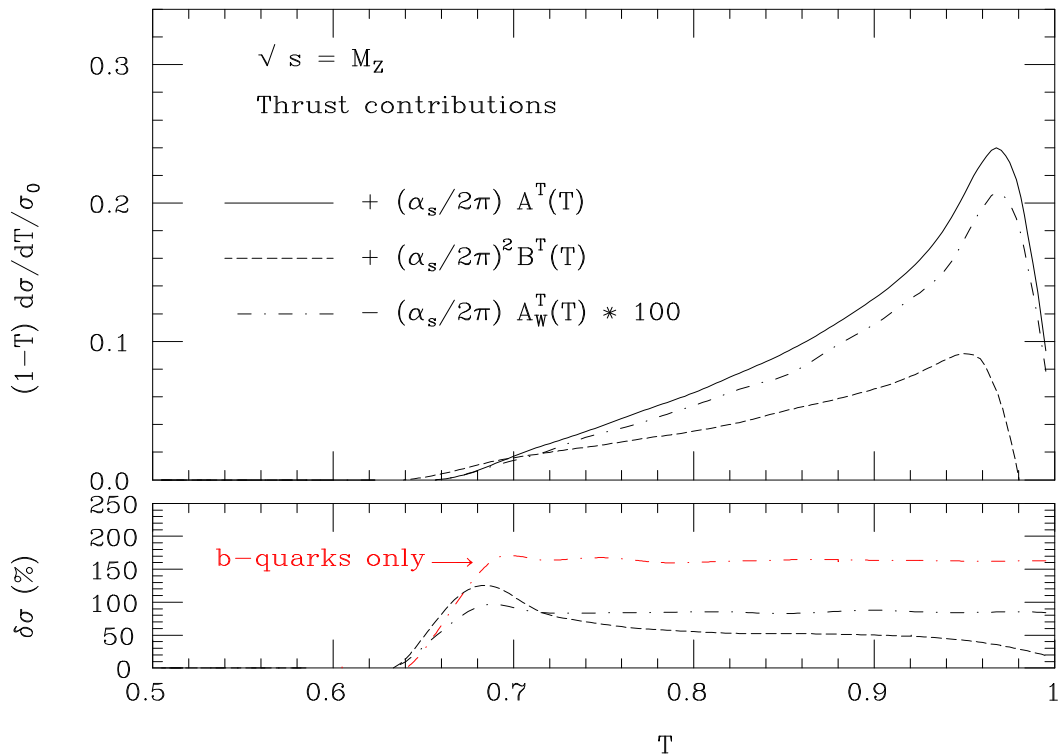}
\end{center}
\vskip -0.5cm
\caption{The LO, NLO-QCD and NLO-W  
contributions to the coefficient functions entering the
integrated Thrust distribution, see eq.~(\ref{T}), for
process (\ref{procj}) (top) and the relative 
size of the two NLO corrections (bottom), at $\sqrt s=M_Z$.
The correction for the case of $b$-quarks only is also
presented, relative to the LO results for process (\ref{procb}).
 (Notice that the
$\sim A_{\mathrm{W}}$ terms
have been plotted with opposite sign and
multiplied by hundred 
for better presentation.)}
\label{fig:thrust}
\end{figure}

We plot the terms $\left(\frac{\as}{2\pi}\right)A^{\Thrust}(\thrust)$,
$\left(\frac{\as}{2\pi}\right)A^{\Thrust}_{\rm{W}}(\thrust)$ and 
$\left(\frac{\as}{2\pi}\right)^2B^{\Thrust}(\thrust)$ in Fig.~\ref{fig:thrust},
always at  $\sqrt s=M_Z$,
alongside the relative rates of the NLO-QCD and NLO-W terms 
with respect to the LO contribution. Here, it can be seen that
the NLO-W effects can reach the level of $-1\%$ or so and that they are
fairly constant for $0.7\lsim {\rm{T}}\lsim 1$. For the case of $b$-quarks
only, similarly to what seen already for the inclusive
rates, the NLO-W corrections are larger, as they can reach the $-1.6\%$ level.

The ability to polarise electron (and possibly, positron) beams joined
with the high luminosity available
render future LCs a privileged environment in which to test the
structure of hadronic samples. As noted earlier,
differential spectra may well carry the distinctive hallmark of some new
and heavy strongly interactive particles (such as squarks
and gluinos in Supersymmetry), whose rest mass is too
large for these to be produced in pairs
as real states but that may enter as virtual 
objects into multi-jet events. 
Similar effects may however also be induced by the NLO-W corrections tackled 
here. Both could well be isolated in one or more of the nine form-factors 
given in eq.~(\ref{FFs}). As already intimated
in the previous Section, $F_7$ to $F_9$ are identically zero at 
LO\footnote{This is strictly true only for massless quarks, as, 
for $m_q\ne 0$,
Ref.~\cite{BDS} has shown that $F_9$ becomes non-zero.}, even prior to 
any integration in $\alpha,\beta$
and/or averaging over the $e^+e^-$ helicities. Besides, $F_7$ would remain
zero unless corrections involve parity-violating interactions whereas
$F_3, F_6$ and $F_9$ would not contribute for unpolarised beams.
As for 
$F_1,... F_6$, we should mention that the NLO-W corrections to the
corresponding tree-level distributions were found to be 
$\lsim1.2(2.0)\%$ for left-(right-)handed 
incoming electrons\footnote{For reasons of space we refrain from 
presenting here the NLO-W dependence of $F_1,...F_6$ in term of $x_1$ 
and $x_2$. The files can be requested from the authors.}.

\begin{figure}
\begin{center}
\epsfig{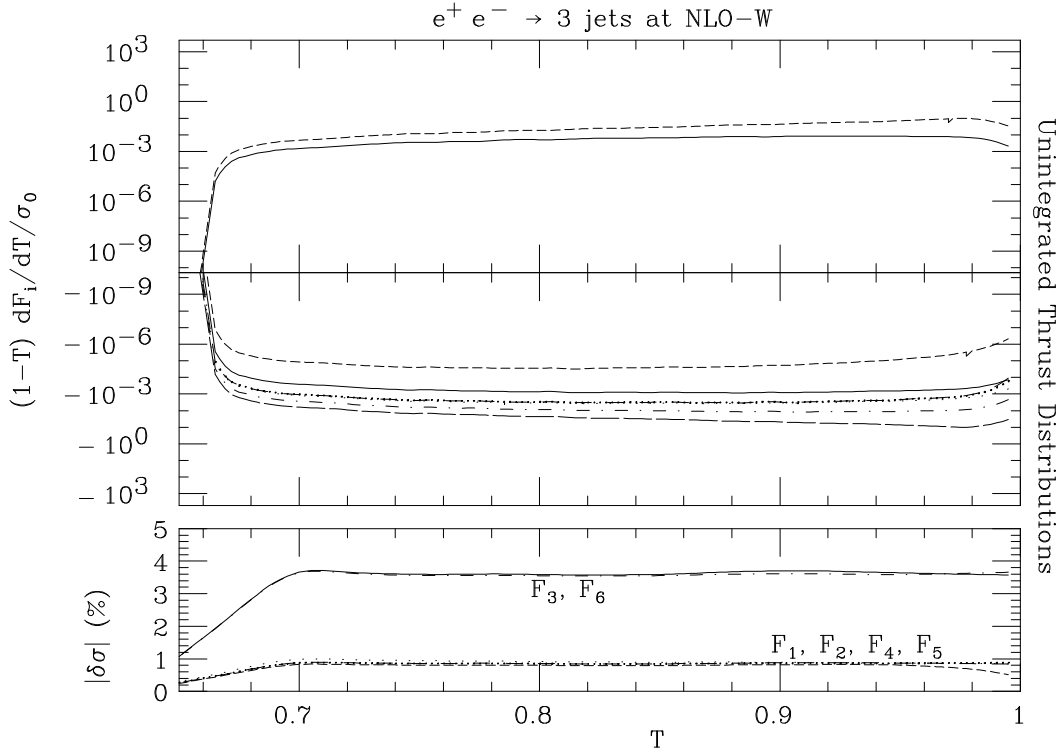}
\caption{The unintegrated Thrust distributions for the nine
component of the cross-section associated to the form-factors
in eq.~(\ref{FFs}) for the NLO-W process (\ref{procj}) (top and middle) 
and the relative size of the
six components which are non-zero at LO (bottom), at $\sqrt s=M_Z$.
Labels are as follows: (top) $F_4$ (solid),
                             $F_3$ (short-dashed);
                    (middle) $F_9$ (solid),
                             $F_8$ (short-dashed),
                             $F_7$ (dotted),
                             $F_6$ (dot-dashed),
                             $F_5$ (dashed),
                             $F_2$ (fine-dotted),
                             $F_1$ (long-dashed);
                    (bottom) $F_6$ (solid),
                             $F_5$ (short-dashed),
                             $F_4$ (dotted),
                             $F_3$ (dot-dashed),
                             $F_2$ (dashed),
                             $F_1$ (fine-dotted).}
\label{fig:allthrust}
\end{center}
\end{figure}

Observables where such effects would immediately be evident are 
what we call the `unintegrated' (or `oriented') Thrust 
distributions associated to each of the form-factors in
eq.~(\ref{angles}) (wherein $S={\rm T}$). In  Fig.~\ref{fig:allthrust},
we present the $F_i\equiv F_i({\rm{T}})$
terms appearing in that expression, 
each divided by $\sigma_0$ (for consistency
with the previous plots), alongside the absolute value of the 
relative size of the NLO-W corrections with respect
to the LO case,
for the form-factors  $F_1,...F_6$, which are non-zero at the Born level.
For the latter, NLO-W corrections can be either positive
or negative, depending on the form-factor being considered, and 
can be as large as $\pm4\%$ or so (in the case of $F_3$ and $F_6$).

\section{Conclusions}
\label{Sec:Conclusions}

On the basis of our numerical findings in the previous Section, we should like
to conclude as follows.

\begin{itemize}
\item At $\sqrt s=M_Z$, the size of the NLO-W corrections to three-jet
rates is
rather small, of order percent or so, hence confirming that determinations
of $\as$ at LEP1 and SLC are stable in this respect and that the
SM background to parity-violating effects possibly induced by new physics
is well under control. In contrast, NLO-W effects 
ought to be included in the case
of future high-luminosity LCs running at the $Z$ pole, such as GigaZ,
where the accuracy of $\as$  measurements from jet rates
is expect to reach the
$0.1\%$ level.

\item Exclusive observables in three-jet events are
 also affected by similar
NLO-W effects: e.g.,  the Thrust distribution, as representative
of the so-called `infrared safe' quantities. The
experimental error expected at LCs in the determination of
$\as$ from such quantities at $\sqrt s=M_Z$ is again
of the order of $0.1\%$ (or even smaller), so that  the
inclusion of NLO-W effects in the corresponding theory predictions 
is then mandatory.

\item Effects from NLO-W corrections are somewhat larger in the case
of $b$-quarks in the final state, in comparison to the case in which
all flavours are included in the hadronic sample, because of the
presence of the top quark in the one-loop virtual contributions. 

\item Since the exploitation of beam polarisation effects will be a
key feature of experimental analyses of hadronic events at future LCs,
we have computed the full differential structure of three-jet processes
in the presence of polarised electrons and positrons, in terms of the energy 
fractions of the two leading jets and of two angles describing the final
state orientation. 
The cross-sections were then parametrised by means of nine independent 
form-factors, the latter presented as a function of Thrust at
fixed angles. Three of these form-factors carry parity-violating
effects which cannot then receive contributions from ordinary QCD.
For the two that are non-zero at LO, the NLO-W 
corrections were found as large as $4\%$. Such higher-order weak effects
should appropriately be subtracted from hadronic samples in the search for 
physics beyond the SM.

\item All our results were presented for the case of the factorisable
NLO-W effects, i.e., for corrections to the initial and final states only.
Whereas these should be sufficient to describe adequately
the phenomenology of three-jet events at LEP1, SLC and GigaZ energies, 
at TeV energy
scales one expects comparable effects due to the non-factorisable corrections,
in which weak gauge bosons connect via one-loop diagrams electrons and 
positrons
to quarks and antiquarks. Their computation is currently in progress and we 
will report on it in due course.

\end{itemize}

\section*{Acknowledgements}

EM and DAR are grateful to the CERN Theory Division and SM to the KEK
Theory Division for hospitality while
part of this work was been carried out. SM and DAR are grateful to John
Ellis for illuminating discussions during the early stages of this project.
This research is supported in part
by a Royal Society Joint Project within
the European Science Exchange Programme (Grant No. IES-14468).

\newpage

\newpage

\vspace{3.0truecm}

\newpage

\newpage

\newpage

\newpage

\begin{thebibliography}{99}

\bibitem{LCs}
K.~Abe {\it et al.}, [The ACFA Linear Collider Working Group],
{\tt hep-ph/0109166};
T.~Abe {\it et al.}, [The American Linear Collider Working Group],
{\tt hep-ex/0106055}; {\tt hep-ex/0106056}; {\tt hep-ex/0106057};
{\tt hep-ex/0106058};
J.A. Aguilar-Saavedra {\it et al.}, [The 
ECFA/DESY LC Physics Working Group],  {\tt hep-ph/0106315};
G. Guignard (editor), [The CLIC Study Team], preprint CERN-2000-008 (2000).

\bibitem{Kuroda:1991wn}
M.~Kuroda, G.~Moultaka and D.~Schildknecht,
\npb{350}{1991}{25};\\
G.~Degrassi and A.~Sirlin,
\prd{46}{1992}{3104};\\
A.~Denner, S.~Dittmaier and R.~Schuster,
\npb{452}{1995}{80};\\
A.~Denner, S.~Dittmaier and T.~Hahn,
\prd{56}{1997}{117};\\
A.~Denner and T.~Hahn,
\npb{525}{1998}{27}.

\bibitem{Beenakker:1993tt}
W.~Beenakker, A.~Denner, S.~Dittmaier, R.~Mertig and T.~Sack,
\npb{410}{1993}{245};
\plb{317}{1993}{622}.

\bibitem{Ciafaloni:1999xg}
P.~Ciafaloni and D.~Comelli,
\plb{446}{1999}{278}
[\hepph{9809321}].

\bibitem{Ciafaloni:2000df}
M.~Ciafaloni, P.~Ciafaloni and D.~Comelli,
\prl{84}{2000}{4810}
[\hepph{0001142}].

\bibitem{Ciafaloni:2001vt}
M.~Ciafaloni, P.~Ciafaloni and D.~Comelli,
{\it Phys. Rev. Lett.} {\bf 87} (2001) 211802 [\hepph{0103315}].

\bibitem{Ciafaloni:2001vu}
M.~Ciafaloni, P.~Ciafaloni and D.~Comelli,
\npb{613}{2001}{382}
[\hepph{0103316}].

\bibitem{Beccaria:2000fk}
M.~Beccaria, P.~Ciafaloni, D.~Comelli, F.~M.~Renard and C.~Verzegnassi,
\prd{61}{2000}{073005}
[\hepph{9906319}].

\bibitem{Beccaria:2000xd}
M.~Beccaria, P.~Ciafaloni, D.~Comelli, F.~M.~Renard and C.~Verzegnassi,
\prd{61}{2000}{011301}
[\hepph{9907389}].


\bibitem{Beccaria:2001jz}
M.~Beccaria, F.~M.~Renard and C.~Verzegnassi,
\prd{63}{2001}{053013}
[\hepph{0010205}].

\bibitem{Beccaria:2001vb}
M.~Beccaria, F.~M.~Renard and C.~Verzegnassi,
\prd{63}{2001}{095010}
[\hepph{0007224}].

\bibitem{Beccaria:2001yf}
M.~Beccaria, F.~M.~Renard and C.~Verzegnassi,
{\it Phys. Rev.} {\bf D 64} (2001) 073008
[\hepph{0103335}].

\bibitem{Denner:2001mn}
A.~Denner,
talk given at the `
International Europhysics Conference on High Energy Physics', 
Budapest, Hungary, July 12-18, 2001
[\hepph{0110155}].

\bibitem{Beccaria:2001an}
M.~Beccaria, S.~Prelovsek, F.~M.~Renard and C.~Verzegnassi,
\prd{64}{2001}{053016}
[\hepph{0104245}].

\bibitem{Denner:2001jv}
A.~Denner and S.~Pozzorini,
\epjc{18}{2001}{461} 
[\hepph{0010201}].

\bibitem{Denner:2001gw}
A.~Denner and S.~Pozzorini,
\epjc{21}{2001}{63}
[\hepph{0104127}].

\bibitem{Pozzorini:2001} 
S.~Pozzorini, {\em Ph.D. Dissertation}, Universit\"at Z\"urich ({2001}).

\bibitem{Accomando:2001fn}
E.~Accomando, A.~Denner and S.~Pozzorini,
{\it Phys. Rev.} {\bf D 65} (2002) 073003
[\hepph{0110114}].

\bibitem{Melles:2000ed}
M.~Melles,
\plb{495}{2000}{81}
[\hepph{0006077}].

\bibitem{Hori:2000tm}
M.~Hori, H.~Kawamura and J.~Kodaira,
\plb{491}{2000}{275}
[\hepph{0007329}].

\bibitem{Beenakker:2000na}
W.~Beenakker and A.~Werthenbach,
{\it Nucl.\ Phys.\ Proc.\ Suppl.}  {\bf 89} (2000) 88
[hep-ph/0006009].

\bibitem{Beenakker:2000kb}
W.~Beenakker and A.~Werthenbach,
\plb{489}{2000}{148}
[\hepph{0005316}].

\bibitem{Beenakker:2001}
W.~Beenakker and A.~Werthenbach, 
{\it Nucl. Phys.} {\bf B 630} (2002) 3
[\hepph{0112030}].

\bibitem{Fadin:2000bq}
V.~S.~Fadin, L.~N.~Lipatov, A.~D.~Martin and M.~Melles,
\prd{61}{2000}{094002} 
[\hepph{9910338}].

\bibitem{Ciafaloni:2000ub}
P.~Ciafaloni and D.~Comelli,
\plb{476}{2000}{49}
[\hepph{9910278}].

\bibitem{Kuhn:2000nn}
J.~H.~K\"uhn, A.~A.~Penin and V.~A.~Smirnov,
\epjc{17}{2000}{97}
[\hepph{9912503}].

\bibitem{Kuhn:2001hz}
J.~H.~K\"uhn, S.~Moch, A.~A.~Penin and V.~A.~Smirnov,
{\it Nucl. Phys.} {\bf B 616} (2001) 286
[\hepph{0106298}].

\bibitem{Melles:2001gw}
M.~Melles,
\prd{63}{2001}{034003}
[\hepph{0004056}].

\bibitem{Melles:2001ia}
M.~Melles,
\prd{64}{2001}{014011}
[\hepph{0012157}].

\bibitem{Melles:2001mr}
M.~Melles,
\prd{64}{2001}{054003}
[\hepph{0102097}].

\bibitem{Melles:2001ye}
M.~Melles,
{\it Phys. Rep.} {\bf 375} (2003) 219. 
[\hepph{0104232}].


\bibitem{Melles:2001dh}
M.~Melles,
{\it Eur. Phys. J.} {\bf C 24} (2002) 193
[\hepph{0108221}].

\bibitem{Layssac:2001ur}
J.~Layssac and F.~M.~Renard,
\prd{64}{2001}{053018}
[\hepph{0104205}].

\bibitem{2jet} See, e.g.:\\
M. Consoli and W. Hollik (conveners), in proceedings of the workshop
`$Z$ Physics at LEP1' (G. Altarelli, R. Kleiss and C.
Verzegnassi, eds.), preprint CERN-89-08, 21 September 1989
(and references therein).

\bibitem{4jet} V.A. Khoze, D.J. Miller, S. Moretti and 
W.J. Stirling, \jhep {07}{1999}{014}.

\bibitem{Dissertori} 
G. Dissertori, talk presented at the `XXXI  International Conference
on High Energy Physics', Amsterdam, 24-31 July 2002, preprint  
{\tt hep-ex/0209070} (and reference therein).
 
\bibitem{Winter} M. Winter, LC Note LC-PHSM-2001-016, February 2001
(and references therein).

\bibitem{ERT} R.K. Ellis, D.A. Ross and A.E. Terrano, \npb{178}{1981}{421}.

\bibitem{VP} G. Passarino and M.J.G. Veltman, {\it Nucl. Phys.}
{\bf B160} (1979) 151. 

\bibitem{FORM} J.A.M. Vermaseren, preprint NIKHEF-00-032
[{\tt math-ph/0010025}].

\bibitem{FeynCalc} J. K{\"u}blbeck, M. B{\"{o}}hm and A. Denner, \cpc
{64}{1991}{165}.

\bibitem{FF1.9} G.J. van Oldenborgh, \cpc {66}{1991}{1}. 

\bibitem{BDS} A. Brandenburg, L. Dixon and Y. Shadmi,
\prd{53}{1996}{1264} [{\tt hep-ph/9505355}].

\bibitem{KS} J.G. K\"{o}rner and G. Schuler, \zpc {26}{1985}{559};
K. Hagiwara, T. Kuruma and Y. Yamada,
{\it Nucl. Phys.} {\bf B 358} (1991) 80.

\bibitem{BMM}A.~Ballestrero, E.~Maina and S.~Moretti,
{\it Phys.\ Lett.}  {\bf B 294} (1992) 425;
{\it Nucl. Phys.} {\bf B 415} (1994) 265 [{\tt hep-ph/9212246}].

\bibitem{bbgNLO}
G.~Rodrigo, A.~Santamaria and M.~Bilenky, 
{\it Phys. Rev. Lett.} {\bf 79} (1997) 193 [{\tt hep-ph/9703358}];
J. Phys. {\bf G 25} (1999) 1593 [{\tt hep-ph/9703360}];
preprint FTUV/98-20, IFIC/98-20 [{\tt hep-ph/9802359}];\\
 G.~Rodrigo, preprint ISBN 84-370-2989-9
[{{\tt hep-ph/9703359}}]; {\em Nucl. Phys. Proc. Suppl.}
  {\bf 54 A} (1997) 60 [{\tt hep-ph/9609213}];\\
W.~Bernreuther, A.~Brandenburg and P.~Uwer, 
{\em Phys. Rev. Lett.} {\bf 79} (1997) 189 [\hepph{9703305}];\\
A.~Brandenburg and P.~Uwer, {\em Nucl. Phys.} {\bf B 515} (1998) 279
[\hepph{9708350}];\\
P.~Nason and C.~Oleari, {\em Phys. Lett.} {\bf B 407} (1997) 57 
[\hepph{9705295}];
{\em Nucl. Phys.} {\bf B 521} (1998) 237 [\hepph{9709360}].

\bibitem{DAR} D.A. Ross, \npb{188}{1981}{109}.

\bibitem{BN} F. Bloch and A. Nordsieck, 
{\it Phys.\ Rev.\ }{\bf 52} (1937) 54.

\bibitem{KLN} T. Kinoshita, {\it J. Math. Phys.\ }{\bf 3} (1962) 650;\\
T.D. Lee and M. Nauenberg, {\it Phys.\ Rev.\ }{\bf 133} (1964) 1549.

\bibitem{EERAD} F.A. Berends, W.T. Giele and H. Kuijf,
{\it Nucl. Phys.} {\bf B 321} (1989) 595; W.T. Giele 
and E.W.N. Glover, \prd{46}{1992}{1980}.

\bibitem{jade}
JADE Collaboration, \zpc {33} {1986} {23};\\ 
S.~Bethke, {\it Habilitation thesis}, preprint LBL 50-208 (1987).

\bibitem{durham}
Yu.L.\ Dokshitzer, contribution cited in the `Report of the 
Hard QCD Working Group', in
Proceedings of the workshop `Jet Studies at LEP and HERA',
       Durham, December 1990, {\it J. Phys} {\bf G 17} (1991) 1537;\\
S.~Catani, Yu.L.~Dokshitzer, M.~Olsson, G.~Turnock and B.R.~Webber,
\plb {269} {1991} {432}.

\bibitem{cambridge} Yu.L.~Dokshitzer, G.D.~Leder, S.~Moretti and
  B.R.~Webber, \jhep {08} {1997} {001}.

\bibitem{schemes} S. Moretti, L. L\"onnblad and T. Sj\"ostrand,
\jhep {08} {1998} {001}.

\bibitem{BKSS}
S.~Bethke, Z.~Kunszt, D.E.~Soper and W.J.~Stirling,
{\it Nucl. Phys.}
 {\bf B 370} ({1992}) {310}; Erratum, preprint {\tt hep-ph/{9803267}}.

\bibitem{mb} 
M. Bilenky, G. Rodrigo and A. Santamaria,
talk given at the IVth International Symposium on 
Radiative Corrections (RADCOR98), Barcelona, Catalonia,
       Spain, 8-12 September 1998, preprint FTUV/98-94, IFIC/98-95
[{\tt  hep-ph/9812433}]; 
contribution to the XXIX International Conference on
       High Energy Physics, Vancouver, Canada, July 1998,
preprint  FTUV/98-79, IFIC/98-80
[{\tt hep-ph/9811465}];\\
M. Bilenky, S. Cabrera, J. Fuster, S. Marti, G. Rodrigo and A. Santamaria,
{\it Phys. Rev.} {\bf D 60} (1999) 114006.

\bibitem{as4}
Z. Nagy and Z. Tr\'ocs\'anyi,
{Phys. Rev.} {\bf D 57} (1998) 5793; {\it ibidem} 
{\bf D 59} (1999) 014020; Erratum,
{\it ibidem} {\bf D 62} (2000) 099902.

\bibitem{KunsztNason} See, e.g.: 
Z. Kunszt and P. Nason,  (conveners), in proceedings of the workshop
`$Z$ Physics at LEP1' (G. Altarelli, R. Kleiss and C.
Verzegnassi, eds.), preprint CERN-89-08, 21 September 1989
(and references therein).

\bibitem{thrust} E. Fahri, \prl{39}{1977}{1587}.

\bibitem{tree5p1}
K.~Hagiwara and D.~Zeppenfeld, {\it Nucl.~Phys.}~{\bf B 313} (1989) 560.

\bibitem{tree5p2}
F.A.~Berends, W.T.~Giele and  H.~Kuijf,  {\it Nucl.~Phys.}~{\bf B 321} 
(1989) 39. 

\bibitem{tree5p3}
N.K.~Falck, D.~Graudenz and G.~Kramer, {\it Nucl.~Phys.}~{\bf B 328} 
(1989) 317.

\bibitem{onel4p1}
Z.~Bern, L.J.~Dixon, D.A.~Kosower and S.~Weinzierl,
{\it Nucl.\ Phys.}  {\bf B 489} (1997) 3
[{\tt hep-ph/9610370}].

\bibitem{onel4p2}
Z.~Bern, L.J.~Dixon and D.A.~Kosower,
{\it Nucl.\ Phys.}  {\bf B 513} (1998) 3
[{\tt hep-ph/9708239}].

\bibitem{onel4p3}
E.W.N.~Glover and D.J.~Miller,
{\it Phys.\ Lett.}  {\bf B 396} (1997) 257
[{\tt hep-ph/9609474}].

\bibitem{onel4p4}
J.M.~Campbell, E.W.N.~Glover and D.J.~Miller,
{\it Phys.\ Lett.}  {\bf B 409} (1997) 503
[{\tt hep-ph/9706297}].

\bibitem{twol3p}
L.W. Garland, T. Gehrmann, E.W.N. Glover, A. Koukoutsakis, and E. Remiddi, 
{\it Nucl. Phys.} {\bf B 627} (2002) 107 [\hepph{0112081}];
{\it Nucl. Phys.} {\bf B 642} (2002) 227 [\hepph{0206067}].
 
\bibitem{ordering} See, e.g.: T. Hebbeker,
{\it Phys. Rep.} {\bf 217} (1992) 69 (and references therein).


\end{thebibliography}
\end{document}